\documentstyle{article}

\renewcommand{\d}{{\rm d}}
\newcommand{\card}{{\rm card}}
\newcommand{\e}{\epsilon}
\newcommand{\Qu}{{\rm Qu}}
\newcommand{\ar}{\longrightarrow}
\newcommand{\n}{\smallskip}
\newcommand{\nn}{\medskip}
\newcommand{\nnn}{\bigskip}
\newcommand{\w}{\omega}
\newcommand{\s}{\sigma}
\renewcommand{\S}{\Sigma}
\newcommand{\la}{\lambda}
\renewcommand{\a}{\alpha}
\begin{document}
\title{Quantum Computers Speed Up Classical with Probability
Zero}
\author{Yuri Ozhigov}
\date{}
\maketitle
{\it  Department of mathematics, Moscow state technological
 University "Stankin", Vadkovsky per. 3a, 101472, Moscow, Russia,
 e-mail: \ y@ oz.msk.ru}
\nnn

{ \bf Abstract}

Let $f$ denote length preserving function on words.
A classical algorithm can be considered as $T$ iterated 
applications  of  black box representing $f$, beginning with input word $x$ 
of length $n$. 

It is proved that if $T=O(2^{\frac{n}{7+\varepsilon}} )$, $\varepsilon >0$, and $f$ is chosen randomly then with
probability 1 every quantum computer requires not less than $T$ 
evaluations of $f$ to obtain the result of classical computation.
It means that the set of
classical algorithms admitting quantum speeding up has probability measure zero.

The second result is that for arbitrary classical time complexity $T$ and $f$
chosen randomly with probability 1 every quantum simulation of classical
computation requires at least $\Omega ( \sqrt {T} )$ evaluations of $f$.

\section{Introduction}

In few recent years the overwhelming majority of studies on quantum algorithms
demonstrated its strength compared with classical ones (look at \cite{BB} ,\cite{DJ} ,
\cite{Sh} ). The most known
advance here is Grover's result about time $O(\sqrt{N} )$ of quantum exhaustive 
search in area of cardinality $N$ (\cite{Gr} ).

However, there exist natural problems for which quantum computer 
can not speed up classical ones. 
Let $\w^*$ denote the set of all words in alphabet $\w$. 
For a length preserving function $f:\ \{ 0,1\}^* \ar\{ 0,1\}^*$ and $x\in\{ 0,1\}^n$  
the result of $k$ 
iterated applications of $f$ is defined by the following induction
$f^{\{ 0\}} (x)=x ,\ f^{\{ k+1\}} (x)=f(f^{\{k\}} (x))$. In the work \cite{Oz97} it
 is proved that the result of 
this computation:

\begin{equation}
x\ar f(x)\ar f(f(x))\ar \ldots\ar \underbrace{f(\ldots f}_T  (x)\ldots )=
f^{\{T\} } (x)
\label{1}
\end{equation}
cannot be obtained by a quantum computer substantially faster than by classical
if $T=O(2^{n/7} )$.

What is the significance of such black box model? The point is that the 
following principle is informal corollary from classical theory of algorithms.

\nn
{\bf Principle of relativization}
{\it Every general method which can be relativized remains valid
 after relativization.}
\nn

Given a code of classical algorithm the only way to obtain the result of
its action on input word $x$ of length $n$ is to run this algorithm on $x$.
In course of computation the code of algorithm can be applied only as 
black box because in general case we can not analyze its interior construction.
Therefore we can assume that a typical classical computation has the form (1) where
a length preserving function $f$ is used as 
oracle. Time complexity of this computation is $T$ in within constant
factor.

The result of \cite{Oz97} was strengthened in the works \cite{FGGS98} and \cite{BBCMW98}
to arbitrary $T$. Namely, both these works proved independently that
every quantum computation of the metafunction PARITY :
$\ \  {\rm Par}\ (g)=\bigoplus
\limits_x g(x)$  of a function $g:\ \{ 0,1\}^n \ar \{ 0,1\}$ requires exactly $2^{n-1}$ 
evaluations of $g$ (half as many as classical). 

The work \cite{BBCMW98} studied computations of metafunctions of the form
$F:\ \{g\} \ar \{ 0,1\}$ where $\{ g\}$ is the set of functions of the form $
g:\ \{ 0,1\}^n \ar\{ 0,1\}$. Specifically, it is proved that if 
$T=o(2^n ) ,\ n\ar\infty$,
then only vanishing part of such metafunctions can be computed exactly with 
$T$ evaluations of $g$ on quantum computer.
The only known way to obtain lower bounds for iterated applications
of black box from lower bounds for metafunctions is computation of PARITY.
The algorithm for computation of ${\rm Par}\ (g)$ can be represented as iterated
application of particular black box which uses $g$ as subroutine. The set
of particular "PARITY"-black boxes have probability measure zero among all
possible black boxes, hence last two works remain the possibility that for
some fairly large part of oracles there exists quantum speeding up of their 
iterations.

In the present work we prove that if $T$ is not very large
then the set
of black boxes whose $T$ iterations admit any quantum speeding up has probability
measure zero.

\nn
\newtheorem{Theorem}{Theorem}
\begin{Theorem}
If $T=O(2^{\frac{n}{7+\varepsilon}} ),\ \varepsilon >0$, then for a black box 
$f$ chosen 
randomly with probability 1 every quantum computation
of $T$ iterations of $f$ requires $T$ evaluations of $f$.

\end{Theorem}
\nn

For arbitrary number $T$ of iterations more weak lower bound for quantum
simulation is established in the following

\nn
\begin{Theorem}
For a black box $f$ chosen randomly with probability 1 every quantum 
computation
of $T$ iterations of $f$ requires $\Omega (\sqrt{T} )$ evaluations of $f$.
\end{Theorem}

\nn

\section{Outline of Quantum Computations}

Oracle quantum computers will be treated here within the framework of approach
proposed by C.Bennett, E.Bernstein, G.Brassard and U.Vazirani in the work
\cite{BBBV} . They considered a quantum Turing machine with oracle as a
model
of quantum computer (for the definitions look also at \cite{BV} ).
 In this paper we use slightly different model of
quantum computer with separated quantum and classical parts, but the
results hold also for the quantum Turing machines.
 We proceed with the exact definitions.

Our quantum query machine consists of two parts: quantum and classical.
\nn

{\bf Quantum part.}
\n

It consists of two infinite tapes: working and query, the finite set
$\cal U$ of unitary transformations which can be easily performed by
the physical devices, and infinite set $F=\bigcup\limits_{n=1}^\infty
F_n$ of unitary transformations called an oracle for the
length preserving function
$f:\ \{ 0,1\}^* \ar\{ 0,1\}^*$, each $F_n$ acts on $2^{2n}$ dimensional
Hilbert space spanned by $\{ 0,1\}^{2n}$ as follows:
$F_n | \bar a,\bar b \rangle = | \bar a ,f(\bar a) \bigoplus \bar b \rangle$,
$\bar a ,\bar b \in\{ 0,1\}^n$, where $\bigoplus$ denotes the bitwise addition
modulo 2.

The cells of tapes are called qubits. Each qubit takes values from the
complex 1-dimensional sphere of radius 1: $\{ z_0 {\bf 0} +z_1 {\bf 1} \
| \ z_1 ,z_2 \in {\tt C}, |z_0 |^2+|z_1 |^2
 =1\}$. Here $\bf 0$ and $\bf 1$ are referred as basic states of qubit
and form the basis of ${\tt C}^2$.

During all the time of computation the both tapes are limited each by two
markers with fixed positions, so that on the working (query) tape only
qubits $v_1 ,v_2 ,\ldots ,v_\tau$ ($v_{\tau+1} ,v_{\tau+2} ,\ldots ,v_{\tau+2n}$)
are available in a computation with time complexity $\tau=\tau(n)$
on input of length $n$. Put $Q=\{ v_1 ,v_2 ,\ldots ,v_{\tau+2n} \}$.
A basic state of quantum part is a function of the form $e:\ Q\ar\{ 0,1\}$.
Such a state can be encoded as $|e(v_1 ) ,e(v_2 ) ,\ldots ,e(v_{\tau+2n})
\rangle$ and naturally identified with the corresponding word in alphabet
$\{ 0,1\}$. Let $K=2^{\tau+2n}$; $\ e_0 ,e_1 ,\ldots ,e_{K-1}$ be all basic
states taken in some fixed order, $\cal H$ be $K$ dimensional Hilbert
space with orthonormal basis $e_0 ,e_1 ,\ldots ,e_{K-1}$. $\cal H$ can
be regarded as tensor product ${\cal H}_1 \bigotimes {\cal H}_2 \bigotimes
\ldots \bigotimes{\cal H}_{\tau+2n}$ of 2 dimensional spaces, where ${\cal H}_i$
is generated by all possible values of $v_i ,\ i=1,2,\ldots ,\tau+2n$.
A (pure) state of quantum part is such an element $x\in\cal H$ that $|x|=1$.

Time evolution of quantum part at hand is determined by two types of
unitary transformations on its states: working and query.
Let a pair $G,U$ be somehow selected, where $G\subset\{ 1,2,\ldots ,\tau+2n\}$,
$U\in\cal U$ is unitary transform on $2^{\card (G)}$ dimensional Hilbert space.

{\it Working transform} $W_{G,U}$ on $\cal H$ has the form $E\bigotimes U'$,
where $U'$ acts as $U$ on $\bigotimes\limits_{i\in G} {\cal H}_i$ in the basis
at hand, $E$ acts as identity on $\bigotimes\limits_{i\notin G} {\cal H}_i$.

{\it Query transform} $\Qu _f$ on $\cal H$ has the form $E\bigotimes F'_n$,
where $F'_n$ acts as $F_n$ on $\bigotimes\limits_{i=\tau+1}^{\tau+2n} {\cal H}_i$ 
and
$E$ acts as identity on $\bigotimes\limits_{i=1}^\tau {\cal H}_i$.

{\it Observation} of the quantum part. If the quantum part is in state
$\chi =\sum\limits_{i=0}^{K-1} \la_i e_i ,$ an observation is a procedure 
which gives
the basic state $e_i$ with probability $|\la _i |^2$.

\nn
{\bf Classical part}.
\n

It consists of two classical tapes: working and query,
which cells are in one-to-one correspondence with the respective qubits of
the quantum tapes and have boundary markers on the corresponding positions.
Every cell of classical tapes contains a letter from some finite
 alphabet $\w$.
Evolution of classical part is determined by the classical Turing machine $M$
with a few heads on both tapes and the set of integrated states of heads:
$\{ q_b ,q_w ,q_q ,q_o ,\ldots\}$. We denote by $h(C)$ the
integrated state of heads for a state $C$ of classical part.

Let $D$ be the set of all states of classical part.

{\it Rule of correspondence } between quantum and classical parts has the form
$R:\ D\ar 2^{\{ 1,2,\ldots ,\tau+2n\} } \times\cal U$, where $\forall C\in D$
$R(C)=\langle G,U\rangle$, $U$ acts on $2^{\card (G)}$ dimensional Hilbert
space so that $U$ depends only on $h(C)$, and
the elements of $G$ are exactly the numbers of those cells on classical tape
which contain the special letter $a_0 \in\w$.

A state of quantum computer at hand is a pair $S=\langle Q(S),C(S)\rangle$
where $Q(S)$ and $C(S)$ are the states of quantum and classical parts respectively.

{\it Computation} on quantum computer. It is a chain of transformations of the
following form:
\begin{equation}
S_0 \ar S_1 \ar \ldots \ar S_\tau ,
\label{2}\end{equation}
where for every $i=0,1,\ldots ,\tau-1$ $C(S_i )\ar C(S_{i+1} )$ is transformation
determined by Turing machine M, and the following properties are fulfilled:

if $h(C(S_i ))=q_w$ then $Q(S_{i+1} )=W_{R(C(S_i ))} (Q(S_i ))$,

if $h(C(S_i ))=q_q$ then $Q(S_{i+1} )=\Qu_f (Q(S_i ))$,

if $h(C(S_i ))=q_b$ then $i=0$, $Q(S_0 )=e_0 ,\ C(S_0 )$ is fixed initial state,
corresponding to input word $a\in \{ 0,1\}^n$,

if $h(C(S_i ))=q_o$ then $i=\tau$,

in other cases $Q(S_{i+1} )=Q(S_i )$.

We say that this quantum computer (QC) computes a function $F(a)$ with probability
$p\geq 2/3$ and time complexity $\tau$ if for the computation (\ref{2}) on every input
$a$ the observation of $S_\tau$ and the following routine procedure fixed
beforehand give $F(a)$ with probability $p$. We always can reach any other value
of probability $p_0 >p$ if fulfill computations repeatedly on the same input
and take the prevailing result. This leads only to a linear slowdown
of computation. There are computations with bounded error probability.
If $p=1$ then we have exact computation.

\section{The Effect of Changes in Oracle on the Result of Quantum
Computation}

For a state $e_j =|s_1 , s_2 , \ldots , s_{\tau+2n} \rangle$ of the quantum part
we denote
the word $s_{\tau+1} s_{\tau+2} \ldots s_{\tau+n}$ by $q(e_j )$.
The state $S$ of QC is called query if $h(C(S))=q_q$. Such a state is querying the
oracle
on all the words $q(e_j )$ with some amplitudes.
 Put ${\cal K} =\{ 0,1,\ldots , K-1\}$. Let $\xi=
Q(S)=\sum\limits_{j\in\cal K} \la_j e_j$.
Given a word $a\in\{ 0,1\}^n$ for a
query state $S$ we define:
$$
\delta_a (\xi )=\sum\limits_{j:\ q(e_j )=a} |\la_j |^2 .
$$

It is the probability that a state $S$ is querying the oracle on the word $a$.
In particular, $\sum\limits_{a\in\{ 0,1\}^n } \delta_a (\xi ) =1$.

Each query state $S$ induces the metric on the set of all oracles if for
length preserving functions $f,g$ we define a distance between them by
$$
\d_S (f,g)= \left( {\sum\limits_{a:\ f(a)\neq g(a)} 
\delta_a (\xi ) }\right)^{1/2} .
$$

\newtheorem{Lemma}{Lemma}
\begin{Lemma} Let $\Qu _f ,\ \Qu_g$ be query transforms on quantum part of QC
corresponding to functions $f,g$; $S$ be a query state. Then
$$
|\Qu_f (S) -\Qu_g (S)|\leq 2\d_S (f,g) .
$$
\end{Lemma}

\n
{\bf Proof}

\n
Put ${\cal L} =\{ j\in{\cal K} \ |\ f(q(e_j ))\neq g(q(e_j ))\}$.
We have: $|\Qu_f (S) -\Qu_g (S) |\leq 2(\sum\limits_{j\in\cal L} 
(|\la_j |)^2 )^{1/2}
\leq2\d_S (f,g) .$ Lemma is proved.
\n

Now we shall consider the classical part of computer as a part of working tape.
Then a state of computer will be a point in $K^2$ dimensional Hilbert space
${\cal H}_1$. We denote such states by $\xi ,\chi$ with indices. All 
transformations
of classical part can be fulfilled reversibly as it is shown by C.Bennett
in the work \cite{Be}. This results in that all transformations in computation
(\ref{2}) will be unitary transforms in ${\cal H}_1$. At last we can
 join sequential steps: $S_i \ar S_{i+1} \ar\ldots\ar S_j$ where $S_i 
\ar S_{i+1}$,
$S_j \ar S_{j+1}$ are two nearest query transforms, in one step. So
the computation on our QC acquires the form

$$
\chi_0 \ar \chi_1 \ar \ldots \ar \chi_t ,
$$
where every passage is the query unitary transform and the following
unitary transform $U_i$ which depends only on $i$:
$\ \chi_i \stackrel{\Qu_f }{\ar} \chi '_i
\stackrel{U_i}{\ar} \chi_{i+1}$. We shall denote $U_i (\Qu_f (\xi ))$
by $V_{i,f} (\xi )$, then $\chi_{i+1} =V_{i,f} (\chi_i ),\ i=0,1,\ldots ,t-1$.
Here $t$ is the number of query transforms (or evaluations of the function $f$)
in the computation at hand.
Put $\d_a (\xi )=\sqrt{\delta_a (\xi )}$.
\n

\begin{Lemma}
If $\chi _0 \ar\chi_1 \ar\ldots\ar\chi_t$ is a computation with oracle for $f$, a
function $g$ differs from $f$ only on one word $a\in\{ 0,1\}^n$ and
$\chi _0 \ar\chi '_1 \ar\ldots\ar\chi '_t$ is a computation on the same QC with a
new oracle for $g$, then
$$
|\chi _t -\chi '_t |\leq 2\sum\limits_{i=0}^{t-1} \d_a (\chi_i ) .
$$
\end{Lemma}
\n

{\it Proof}
\n

Induction on $t$. Basis is evident. Step.
In view of that $V_{t-1,g}$ is unitary, Lemma 1 and
inductive hypothesis,
we have
$$
\begin{array}{l}
|\chi_t -\chi '_t |=|V_{t-1,f} (\chi_{t-1} )-V_{t-1,g} 
(\chi ' _{t-1} )|\leq \\
|V_{t-1,f} (\chi _{t-1} )-V_{t-1,g} (\chi_{t-1} )|+|V_{t-1,g} 
(\chi_{t-1} )-V_{t-1,g}
(\chi '_{t-1} )| \leq \\
2\d_a (\chi _{t-1} ) +|\chi _{t-1} -\chi '_{t-1} |=
2\d_a (\chi _{t-1} ) +
2\sum\limits_{i=0}^{t-2}\d_a (\chi_i )=
2\sum\limits_{i=0}^{t-1}\d_a
(\chi_i ) .
\end{array}
$$
Lemma is proved.

\section{Basics of Probabilistic Method}

To analyze black boxes chosen randomly with some probability
we need some notions of probability theory.

Given a set $\cal N$ we say that some set $\S\subseteq 2^{\cal N}$ of 
its
 subsets is $\s$-algebra  (algebra) on $\cal N$ iff $\ \ \emptyset ,{\cal N} 
\in \S\ $
and $\S$ is closed with regard to operations of subtractions: $A\setminus B$
and denumerable (finite)  joins and intersections: 
$\bigcup\limits_{i=0}^\infty A_i ,\
\bigcap\limits_{i=0}^\infty A_i $.
Elements of $\S$ are called events. 

A probability measure on $\S$ is such a real function on events $P:\ \S\ar 
[0,1]$
that $P(\emptyset )=0,\ P({\cal N} )=1,$ and for every list $\{ A_i \}$
of mutually exclusive events the following axiom of additivity takes place.
$$
P\left( \bigcup\limits_{i=0}^\infty A_i  \right) =\sum\limits_{i=0}^\infty 
P(A_i ).
$$

The minimal $\s$-algebra containing a given algebra $S\subseteq
2^{{\cal N}}$ is denoted by $\S (S)$. 
Every probability measure on algebra $S$ can be extended to the probability 
measure on $\S (S)$.
We shall denote it by the same letter $P$.

Let $M_n$ denotes the set of all mappings $g:\ \{ 0,1\}^n \ar \{ 0,1\}^n$.
Put $\card(M_n ) =v_n .$ We have $v_n =2^{n2^n}$.
Let $F$ be the set of all oracles. An element of $F$ is length preserving 
function $f:\ \{ 0,1\}^* \ar \{ 0,1\}^*$, which may be regarded as a list 
$g_1 ,g_2 ,\ldots $ of the functions $g_i \in M_i$. We are going to define 
the probability measure distributed uniformly on oracles. For any fixed
$g_i \in M_i \ \ i=1,2,\ldots ,n$ put 
$A(g_1 ,g_2 ,\ldots ,g_n ) =\{ f\ |\ f=(g_1 ,g_2 ,
\ldots , g_n , \ldots  )\}$ and define $P(A( g_1 ,\dots ,g_n ))=
(v_1 v_2 
\ldots v_n )^{-1}$. It is readily seen that axiom of additivity is satisfied
for the extension of $P$ to the minimal algebra $S$ containing
all $A(g_1 ,\ldots ,g_n )$ for all $n$ and $g_1 ,g_2 ,\ldots , 
g_n $ and hence to the probability measure on $\S (S).$
\nn

{\bf Definition} The probability measure on oracles distributed uniformly
is the probability $P$ on $\s$-algebra $\S =\S (S)$.
\n

{\bf Example} Given $n$ and two words $x,y\in\{ 0,1\}^n .$ Then the probability of that
 $f(x)=y$ is $P(B_{xy} )$ where $B_{xy} =\{ f\ |\ f(x)=y\} .$
This probability is $2^{-n}$. 
\nn

For events $A,B\in \S ,\ P(B)\neq 0$ the conditional probability is defined by 
$P(A \ |\ B)=P(A\cap B )/P(B)$. Full group of events for $A$ is such set 
$F_1 ,F_2 ,\ldots
,F_m$ of events with nonzero probabilities that $F_i \cap F_j =\emptyset$ for $i\neq j$ and 
$A\in \bigcup\limits_{i=1}^m F_i$.
In this case $P(A)=\sum\limits_{i=1}^m P(A\ |\ F_i )P(F_i )$
( the folmula of full probability).

 \section{Impossibility of Quantum Speeding Up for the Bulk of Short Computations}

{\it Proof of Theorem 1}
\nn

Let $t(n) ,T(n)$ be integer functions, $T=O(2^{\frac{n}{7+\varepsilon }} ) ,
\ \varepsilon >0$, $C$ be quantum computer.
Denote by $S(C,n,t,T)$ the set of such functions $f \in M_n$ that $C$ 
computes $f^{\{ T\} } (\bar 0 )$ using no more than $t$ evaluations 
of $f$, where $\bar 0$ is the word of zeroes.

\begin{Lemma}
For every quantum computer $C$ and $\e >0$ there exists such number $n$
that $P(S(C,n,T-1 ,T))<\e$.
\end{Lemma}

{\it Proof of Lemma 3} 
\nn

To prove Lemma 3 we need some technical propositions.
Put $\a =5+\frac{\varepsilon}{2}$.
Fix integer $n$.

Now we shall define the lists of the form 
$\zeta_i =\langle \xi_i ,f_i ,{\cal T}_i
, x_i \rangle$ where $\xi_i$ is a state from ${\cal H}_1$, $|\xi_i |=1$,
$f_i \in M_n$,
$x_i \in {\cal T}_i \subseteq \{ 0,1\}^n$ by the following induction on $i$.

\n
{\bf Definition }

Basis: $i=0$. Put $\xi_0 =\chi_0$, let $f_0 \in M_n$ be chosen randomly, $
 x_0 =\bar 0$, ${\cal T}_0 =\{ 0,1\}^n$.

Step. Put
$$
\begin{array}{l}
\xi_{i+1} =V_{i,f_i} (\xi_i ) , \\
{\cal T}_{i+1} ={\cal T}_i \cap R_i ,\ \ R_i=\{a\ |\ \delta_a (\xi_{i+1} )<
\frac{1}{T^\a } \} ,
\end{array}
$$
We define $x_{i+1}$ as randomly chosen element of ${\cal T}_{i+1}$ and
put
$$
f_{i+1} =\left\{
\begin{array}{l}
f_i (x),\ \ \mbox{if }\ x\neq x_i ,\\
x_{i+1} ,\ \ \mbox{if }\ x=x_i .
\end{array}
\right.
$$
\n

The lists of the form $\zeta_i$ are not defined uniquely and we denote
the set of all such lists $\zeta_i$ by $D_i ,\ i=1,2,\ldots$.
Let $N_i$ be the set of such functions $f_i \in M_n$ that there exist
$\xi_i ,{\cal T}_i ,x_i$ such that
$\langle \xi_i ,f_i ,{\cal T}_i ,x_i \rangle
\in D_i$.

\n
\newtheorem{Proposition}{Proposition}
\begin{Proposition}
If $i\leq T$, $n\ar\infty$, then
$$
P(N_i )=1-O\left( \frac{T^{\a+1} i}{2^n } \right) .
$$
\end{Proposition}

\n
{\it Proof of Proposition 1}
\n

Induction on $i$. Basis follows from the definition of
$\zeta_0$. Step. Given some list
$\zeta_i =\langle \xi_i ,f_i  ,{\cal T}_i ,x_i \rangle$, in the passage
to $\zeta_{i+1}$ the only arbitrary choice is the choice of $x_{i+1}$. 
This choice
can be done correctly with probability $\frac{2^n -T^{\a +1}}{2^n}$, because
$\card ({\cal T}_i ) >2^n -T^\a i \geq 2^n -T^{\a +1}$.
Hence in view of inductive hypothesis the resulting probability is
$\left(1-O\left( \frac{T^{\a+1} i}{2^n } \right) \right) 
\left( 1-\frac{T^{\a+1}}{2^n} 
\right)$ $=\left(1-O\left(\frac{T^{\a+1} (i+1)}{2^n} \right)\right)$ 
with the same constant.
Proposition 1 is proved.
\n

Now turn to the proof of Lemma 3. Let in what follows
$t=T-1$. Given lists $\zeta_i$,
we introduce the following notations: $V_i =V_{i,f_t} ,\ V_i^* =V_{i,f_i}$.
Let the unitary operator $V^i$ be introduced by the following induction:
$V^0 (x)=V_0 (x), \ V^i (x) =V_i (V^{i-1} (x))$, and the
unitary operator $\tilde V_i$ be defined
by $\tilde V_0 =V_0^* ,$ $\tilde V_i (x) =V_i^* (\tilde V_{i-1} (x))$.
Then $\xi_{i+1} =\tilde V_i (\xi_0 )$.

Put $\xi '_0 =\xi_0 ,$  $ \xi '_{i+1} =V^i (\xi _0 ) ,$
$\ \partial _i =|\xi_i
-\xi '_i | ,$  $ \Delta_i =|V_i^* (\xi_i )-V_i (\xi_i )|$. 
It follows from the
definition that $f_i$ differs from $f_t$ at most on the set $X_i =
\{ x_i ,x_{i+1} ,\ldots ,x_{t-1} \}$ where $\forall a\in X_i$
$\delta_a (\xi_i ) <\frac{1}{T^\a}$. Consequently, applying Lemma 1
we obtain

\begin{equation}
\Delta_i \leq \frac{2t^{1/2}}{T^{\a /2}} .
\label{3}
\end{equation}
\n

\begin{Proposition}
$\ \ \ \ \partial_i \leq\sum\limits_{k<i} \Delta _k .$
\end{Proposition}
\n

{\it Proof}
\n

Induction on $i$. Basis follows from the definitions. Step:
$$
\begin{array}{l}
\partial_{i+1} =|\tilde V_i (\xi_0 ) -V^i (\xi_0 )|=|V_i^* (\tilde V_{i-1}
(\xi_0 ))-V_i (V^{i-1} (\xi_0 ))|\leq \\
\leq |V_i^* (\xi_i )-V_i (\xi_i )|+|V_i (\xi_i )-V_i (\xi '_i )|=
\Delta_i +\partial_i .
\end{array}
$$
Applying the inductive hypothesis we complete the proof. 

Thus in view of (3) Proposition 2 gives
\begin{equation}
\forall i=1,\ldots ,t\ \ \ \partial_i \leq\frac{2it^{1/2}}{T^{\a /2}} .
\label{4}
\end{equation}

It follows from the definition of the functions $f_i$ that 
$\forall i\leq t \ \
\delta_{x_t} (\xi _i )<\frac{1}{T^\a } .$ Taking into account inequality 
(\ref{4} ),
we conclude that for $x=x_t$

$
\d_x (\xi_i -\xi '_i )\leq\frac{2it^{1/2}}{T^{\a /2}},$
$\d_x (\xi _i )<\frac{1}{T^{ \a /2}},$
$\d_x (\xi _i ' )\leq\d_x (\xi_i -\xi '_i )+
\d_x (\xi _i ).$

Hence we have
\begin{equation}
\d_x (\xi_i ')\leq
\frac{3t^{3/2}}{T^{\a /2}}. \label{5}
\end{equation}

Now consider some oracle  $f_{t+1} =f_T$. 
 If
$\xi_0 \ar\xi_1 '' \ar\ldots \ar\xi_t ''$ is the computation of
$f_{t+1}^{\{ T\}} (\bar 0 )$ on our QC with oracle for $f_{t+1}$,
 then  Lemma 2 and inequality (\ref{5}) give
$$
|\xi '_t -\xi ''_t |<2\sum\limits_{i\leq t} \d_x (\xi_i ' )
\leq\frac{6t^{5/2}}{T^{\a /2}} <\gamma (n)
$$
for $\a =5+\frac{\varepsilon}{2}$, where $\gamma (n)$ can be made arbitrary 
small for appropriate
$n$. Hence, observations of states $\xi '_t$ and $\xi ''_t$ give
the same results with closed probabilities. Then if our computer
does computes $f_{t+1}^{\{ T\}} (\bar 0 ) =a$, then amplitudes
of basic states in $\xi '_t$ must concentrate on only one
unique basic state corresponding to $a$.

Let $P( \mbox{not}\ |\ f_t )$ be the probability to choose an oracle of the
form $f_T$ such that $f_T^{\{ T\}} (\bar 0 )
\neq f_{T-1}^{\{ T\} } (\bar 0 )$
given $f_t$.
In view of the definition of computation it is the probability of that with
a given choice of $f_t$ our computer does not compute
$f_T^{\{ T\} } (\bar 0 )$
correctly. We have
$P( \mbox{not}\ |\ f_t ) =\frac{2^n -T^{\a +1}}{2^n} \ar 1\
(n\ar \infty)$
for every choice of $f_t$, because there are at least $2^n -T^{\a +1}$ appropriate
possibilities for the choice of $x_{t+1}$.
Furter, let $\tilde p$ be the probability to choose an oracle $f_T$ such that
 our computer does not
compute $f_T^{\{ T\}} (\bar 0 )$ correctly. With the formula of full
probability and Proposition 1 we have
$$
\tilde p =\sum\limits_{f_t} P( \mbox{not}\ |\ f_t )p(f_t )
=\frac{2^n -T^{\a +1 }}{2^n}
\left( 1-O\left(\frac{T^{\a+2}}{2^n} \right)\right) \ar 1\ (n\ar\infty ).
$$

At last the probability $p_{{\rm not}}$ to choose oracle $f$ such that 
$f^{\{ T\}} (\bar 0 )$
is not computed on computer at hand will be
$p_{{\rm not}} \geq\tilde p$, then $p_{{\rm not}} \ar 1\ (n\ar\infty )$.

Lemma 3 is proved.

\nn

Now turn to the proof of Theorem 1. Let $C_1 , C_2 ,\ldots $ be all 
quantum query machines taken in some fixed order, $R(C,n,t,T)$
denote the set of all functions $g_n \in M_n$ such that 
$C$ does not compute $g_n^{\{ T\} } (\bar 0 )$ using no more than $t$ 
evaluations 
of $g_n$. Take arbitrary $\e >0$. Applying Lemma 3 find for every 
$i=1,2,\ldots $
such number $n_i$ that $P(R(C_i ,n_i ,T-1 ,T))>1-\e 2^{-i}$. Furter,
if ${\cal N}$ denotes the set of oracles which $T$ iterated applications
do not admit quantum speeding up, we have 
$\bigcap\limits_{i=1}^\infty R(C_i ,n_i ,
T-1,T) \subseteq {\cal N} $. For the complementary set 
$\bar{\cal N} \subseteq
\bigcup\limits_{i=1}^\infty S(C_i ,n_i ,T-1,T)$. By axiom of additivity
$P(\bar{\cal N} )\leq\sum\limits_{i=1}^\infty P(S(C_i ,n_i ,T-1,T))$
$=\sum\limits_{i=1}^\infty \e 2^{-i} =\e$.
Theorem 1 is proved.

\section{Lower Bound for Quantum Simulation in General Case}

{\it Proof of Theorem 2}
\n

As in the previous section it would suffice to prove the following 
\n

\begin{Lemma}
For every quantum query machine $C$, $\e >0$ and functions 
$t(n),T(n):\ \ t^2 =o(T) 
\ (n\ar\infty )$ there exists integer $n$ such that 
$P(S(C,n,t,T))<\e$.
\end{Lemma}
\n

{\it Proof of Lemma 4}
\n

In our notations for randomly chosen oracle $f$ and number $n$ put 
$f^k =f^{\{k\}} 
(\bar 0 )$, $k=0,1,\ldots ,T$. Define matrix $(a_{ij} )$ by
$a_{ij} =\delta_{f^j} (\chi_i ) ,\ \ i=0,1,\ldots ,t ;\ j=0,1,\ldots ,T.$

We have for every $i=0,\ldots ,t\ \ \sum\limits_{j=0}^T a_{ij}
\leq 1$, consequently $t\geq\sum\limits_{i=0}^t \sum\limits_{j=0}^T a_{ij} =
\sum\limits_{j=0}^T \sum\limits_{i=0}^t a_{ij}$ and there exists such
$\tau \in\{ 0,1,\ldots ,T\}$ that 
$\sum\limits_{i=0}^t a_{i\tau} \leq\frac{t}{T}$.

Changing arbitrarily the value of $f$ only on the word $f^\tau$ we obtain a new function
$g$ where $g ^{\{T\}} (\bar 0 ) \neq f^{\{ T\}} (\bar 0 )$ with probability 
$p_n \ar 1\ 
(n\ar\infty )$. Let $\chi_0 \ar\chi '_1 \ar\ldots \ar\chi '_t$ be computation on QC with oracle
for $g$. Then for such choice of $g$ with probability $p_n$ we have 
$|\chi_t -\chi '_t |\geq 1/4$ if $f\in S(C,n,t,T)$. 

On the other hand Lemma 2 gives
$
|\chi_t -\chi '_t |\leq 2\sum\limits_{i=0}^t \sqrt{a_{i\tau}} \leq2\sqrt{t
\sum a_{i\tau}} \leq 2t/T^{1/2} <\gamma (n)\ar 0\ (n\ar\infty )$.
Then by the definition of computation with probability 
$\tilde p_n \ar 1 \ (n\ar\infty )$
$\ \ g^{\{ T\}} (\bar 0 )$ is not computed by quantum computer at hand.
Lemma 4 is proved. Theorem 2 is derived from Lemma 4 just as Theorem 1 from Lemma 3.
Theorem 2 is proved.

\newpage
\section{Acknowledgements}
I am grateful to Charles H. Bennett who clarified for me some details of the work
\cite{BBBV}, to Peter Hoyer for useful discussion, to
Lov K. Grover for his attention to my work and to Richard Cleve and Edward Farhi
for the information about the works  \cite{BBCMW98} and \cite{FGGS98}.
I also thank principal of "Stankin" Yuri Solomentsev for the financial support of my work.

\end{document}